\documentclass[preprint,nofootinbib,nobibnotes,groupaddress,preprintnumbers,aps,prd, twocolumnn]{revtex4}
\usepackage{bm,epsfig,mathrsfs,amsmath,amssymb,graphicx}
\usepackage{epsfig,amsmath,graphicx,amssymb}
\usepackage{graphicx,dcolumn,bm}
\usepackage[toc,title,titletoc,header]{appendix}
\def\dbar{{\mathchar'26\mkern-12mu d}}

\begin{document}

\title{Quantum-mechanical engine models and   their efficiencies }
\author{Jianhui Wang$^{1,2}$} \email{wangjianhui@ncu.edu.cn}
\author{Yongli Ma$^2$}
\author{Jizhou He$^1$}
\affiliation{$^1\,$ Department of
Physics, Nanchang University, Nanchang 330031, China \\
$^2\,$ State Key Laboratory of Surface Physics and Department of
Physics, Fudan University, Shanghai 200433, China}

\begin{abstract}
Based on quantum thermodynamic processes, we make a
quantum-mechanical (QM) extension of the typical heat engine cycles,
such as the Carnot, Brayton, Otto, and Diesel cycles, etc. The
temperature is not included in these QM engine cycles, as lies in
the fact that the concept of energy is well-defined in quantum
mechanics, temperature \emph{a priori} is not. These QM  engine
cycles are implemented by an ideal or interacting system with an
arbitrary number of particles confined in an arbitrary power-law
trap. As a result, a relation between the quantum adiabatic exponent
and trap exponent is found. The efficiency of a given QM  engine
cycle is similar to that of its classical counterpart, thereby
identifying the universality of the efficiency.


PACS number(s): 05.70.-a, 03.65. -w, 51.30+i

\end{abstract}

\maketitle
\date{\today}
The current activities in quantum thermodynamics \cite{Gem09} focus
on quantum heat engines \cite{Scu03,Quan07,Quan09,Fia12,Kim11,
Gev92, Ben00,Abe12,pre1104,pre12III, pre1204,Lu12,Wu06,
Rez06,Fel00,pre1203, Huang12, Per98} or refrigerators
\cite{Cle12,He02}, work-extraction processes \cite{Kie04,All08,
Hul12} from quantum systems, and positive work conditions
\cite{Quan05}. Among all the studies about quantum thermodynamics, a
central concern is to make a quantum extension
\cite{Quan07,Quan09,Ben00} of classical thermodynamic processes and
cycles. As in classical thermodynamics, for quantum thermodynamics
there are some basic thermodynamic processes: isothermal, adiabatic,
isochoric, isobaric, and isoenergetic processes. These processes
correspond to constant temperature, entropy, volume, pressure, and
energy, respectively. They can be used to construct all kinds of
thermodynamic cycles, such as the Carnot, Brayton,  Otto, Diesel,
Ericsson, and Stirling cycles, etc.  Because of quantum features of
the working substance, unusual and exotic behaviors have been found
in quantum heat engines.  A prominent example  is a quantum heat
engine which may use an isolated finite system as its working
substance to produce work \cite{Fia12}.  In an isolated finite
system \cite{Hui11} as well as quantum mechanics,  the concept of
energy (rather than temperature) is well-defined.  Recently, a
quantum-mechanical (QM) Carnot cycle working between two energy baths
instead of heat baths has been generalized and studied intensively
\cite{Ben02, Abe12, pre12III} since it was first proposed by  Bender
et al \cite{Ben00}.
  Nevertheless,  little attention was
paid to such a QM  generalization of the remaining classical
thermodynamic processes and cycles until most recently
\cite{pre1204}.

In this paper, we  study various quantum thermodynamic processes and
their related quantum thermodynamic cycles. We begin our analysis
with the definitions of quantum thermodynamic processes, including
isoenergetic, isobaric, and isochoric processes, and with
clarification of  how to achieve these processes. The generalization
of these  processes allows us to study an arbitrary quantum
thermodynamic cycle constructed by any four  of these  processes. We
discuss various quantum thermodynamic cycles, such as the QM Carnot,
Brayton, Otto,  and Diesel cycles, etc., and compare their
properties with their classical counterparts. From these
comparisons, the universality of the efficiency is identified for a
given cycle.

{\emph{Various thermodynamic processes for a QM system.}} For a QM
system, the expectation value of the system Hamiltonian is given by
$ E=\langle\psi|{H}|\psi\rangle=\sum_n p_n {\varepsilon_n}, $ where
$\varepsilon_n$ is the single-particle energy spectrum and $p_n$ is
the mean occupation probability of the $n$th eigenstate, with
$n=1,2,3,\cdots$. The derivation of $E$ leads to the first law of
quantum thermodynamics \cite{Quan09,Kim11,pre1203}: $ d E=\dbar
Q+\dbar W=\sum_{n}\varepsilon_ndp_n+\sum_{n}p_nd\varepsilon_n$,
where $\dbar Q=\sum_{n}\varepsilon_ndp_n$ and  $\dbar
W=\sum_{n}p_nd\varepsilon_n$ are associated with the energy exchange
and work done, respectively. That is, energy exchange between a QM
system and its surroundings is induced by transitions between
quantum states of the system,  in which whether temperature (heat
bath) is included or not, while work is performed due to variation
of energy spectrum with fixed occupation probabilities. As in a
classical system where the generalized force $Y_n$, conjugate to the
generalized coordinate $y_n$, is defined by $Y_n=-\frac{\dbar
W}{d{y_n}}$, the force for a quantum system can be defined as
\begin{equation}
F=-\frac{\dbar W}{d L}=-\sum_n {p_n}\frac{d \varepsilon_n}{d L}.
 \label{ffdl}\end{equation}
Here $L$ as the generalized coordinate corresponds to the force $F$,
which is, in fact, the pressure of the quantum system. 

Without loss of generality, we consider a quantum system which
consists of an arbitrary number of ideal or interacting particles
confined in an arbitrary power-law trap. A one-dimensional power-law
trap can be parameterized by a single-particle energy spectrum of
the form \cite{ epjd, Wil97,pre1104}
\begin{equation}
\varepsilon_{n}=\hbar\omega n^{\sigma}=\hbar\lambda
L^{-\theta}n^\sigma,
 \label{enma}\end{equation}
where $n$ is a positive integer quantum number, and $\sigma$ is an
index of the single-particle energy spectrum. Here we have used the
relation $\omega =\lambda L^{-\theta}$, where $\lambda$ is a
constant for a given potential, and $\theta$ is trap exponent
\cite{wpra09} depending on the trapping potential \cite{pre1104}.
Note that the energy spectrum (\ref{enma}) can also be used to
depict the other physical systems, such as a harmonic system
\cite{Gev92}, a spin-$1/2$ system \cite{Huang12, Gev92}, and a
single-mode radiation field in a cavity \cite{pre1104}, etc. It
follows, on substitution of Eq. (\ref{enma}) into Eq. (\ref{ffdl}),
that the force acting on the potential wall is
\begin{equation}
F=\theta\sum_n p_n\frac{\varepsilon_n}L,
 \label{flnl}
\end{equation}
from which,   the internal energy  $E_i$ of the system at any
instant $i$  can be derived as
\begin{equation}
E_i=\sum_n p_n \varepsilon_n=L_i F_i/\theta . \label{uifi}
\end{equation}

 To realize an isoenergetic process  \cite{pre12III},  the system
exchanges energy with an energy bath
  so that the work done by the external
parameter $\lambda$, on which the Hamiltonian of the quantum system
depends parametrically, can be precisely counterbalanced.  In the
isoenergetic process, the quantum system evolves from an initial
state $|\psi(0)\rangle$ to a final state $|\psi (t)\rangle$ through
a unitary evolution.
One possible way of achieving this is to demand
the constancy of the expectation of the Hamiltonian in such a way
that $\frac{d {H}}{d t}=[{H}(t),{H}(t')]+\frac{\partial
{H}}{\partial t}=[{H}(t),{H}(t')]+\frac{\partial {H}}{\partial
\lambda}\frac{\partial \lambda}{\partial t}=0$,  where $0\leq
t<t'\leq t_{ie}$ with $t_{ie}$ being the time required for
completing the isoenergetic process.

Both the system volume $L$ and the occupation probabilities $p_n$
change in the isoenergetic process, and the system exchanges energy
with the energy bath in order for the system energy $E(L)$ to be
kept constant.  So the energy exchange is a form of heat exchange by
definition, and the external energy baths play the role of heat
baths in conventional heat engines. According to the first law of
thermodynamics, energy absorbed by the system in an isoenergetic
process, with constant energy $E_{if}$, can be determined according
to
\begin{equation}
Q_{if}=\int_{L_i}^{L_f}F_{if}dL=\theta E_{if}\ln{(L_f/L_i)},
\label{qili}
\end{equation}
where we have used  $Q_{if}=W_{if}$ in the isoenergetic process.

For  a quantum isobaric process with constant pressure, the time
scale of relaxation of the system with the heat bath should be much
smaller than that of the variation of the system volume \cite{Lin76,
Quan09}. If a classical or a quantum system coupled to a heat bath
undergoes an isobaric process, we must carefully control the
temperature of the system as well as the temperature of the heat
bath under some conditions that sensitively depend on the systems
\cite{Quan09}, when we change the volume of the system. This is not
the case in the absence of a heat bath. In contrast, we can see from
Eq. (\ref{flnl}) that $LF(L)/E=\texttt{cons},$ which can be regarded
as the equation of state for the system under consideration. The
energy of the system undergoing a quantum isobaric process only
needs to be controlled in such a way that $E\propto L$, which is
independent of the form of the trapping potential. Similar to a
quantum isoenergetic process, the controlled parameter in a quantum
isobaric process is switched from $\lambda(0)$ to $\lambda(t_{ib})$
in a period $t_{{ib}}$. One possible way to achieve the
constant-pressure process is that the pressure $F(\lambda)$, other
than the Hamiltonian $H$, should satisfy the condition: $\frac{d
{F}}{d t}=[{F}(t),{H}(t')]+\frac{\partial {F}}{\partial
t}=[{F}(t),{H}(t')]+\frac{\partial {F}}{\partial
\lambda}\frac{\partial \lambda}{\partial t}=0,$ where $0\leq
t<t'\leq t_{ib}$.

 For an isobaric expansion
$i\rightarrow f$, the energy transferred to the system not only
produces work $W_{if}=(L_f-L_i) F_{if}$ but also changes the
 energy of the system $\Delta E=E_f-E_i$.  The energy absorbed
by the system in the isobaric expansion, $Q_{if}$, is therefore
obtained by use of Eq.  (\ref{uifi}),
\begin{equation}
Q_{if}=E_f-E_i+W_{if}=\frac{\theta+1}{\theta}{(L_f-L_i)}F_{if}.
 \label{qiif}\end{equation}

 An isochoric process is one in which the volume $L$
is held constant, meaning that, while no work is done by the system,
energy as a form of heat is exchanged between the working substance
and the energy bath.  
The transitions between quantum states as well as the variation of
occupation probabilities in an atomic system are achieved when the
system interacts with an external field \cite{pre1204}, which can be
regarded as a good example of a quantum isobaric process without
introduction of temperature. The condition that the working
substance should reach thermal equilibrium with the heat bath at the
end of a conventional quantum as well as classical isochoric process
\cite{Quan07} is therefore no longer required to be fulfilled in
such a quantum isochoric process. Energy quantity absorbed by the
system during a quantum  isochoric process $i\rightarrow f$ is given
by
\begin{equation}
Q_{if}=E_f-E_i.
 \label{qiui}\end{equation}

Although the conditions  that realize the quantum isobaric and
isochoric processes without inclusion of temperature are different
from corresponding ones with introduction of temperature, the heat
exchanges between the system and its surroundings are easily proved
to be still given by Eqs. (\ref{qiif}) and (\ref{qiui}) in the
presence of a heat bath.

 The quantum adiabatic process has been extensively clarified in many
references \cite{Quan09, Foc28, pre12III, arx12} ever since the
birth of quantum mechanics. A quantum adiabatic process must proceed
at  a very slow speed so that the time scale of the change of the
system state must be larger than that of the dynamical one,
$\sim\hbar/E$ \cite{Foc28, Abe12, pre12III} and thus the generic
quantum adiabatic condition \cite{Foc28} is satisfied. The
occupation probabilities remain unchanged, $dp_n=0$, which, together
with the relation $\dbar Q=\sum_n \varepsilon_n d p_n$, means that
there is no heat exchange in a quantum adiabatic process.

Given a trapping potential, the index $\sigma$ of the energy
spectrum and the trap exponent $\theta$ are  fixed. Thus, for a
quantum system undergoing an adiabatic process, we have
\begin{equation}
L^{\theta+1}F=\texttt{cons}.
 \label{ltns}\end{equation}
Through comparison with $ L^\gamma F=\texttt{cons}$ for the
classical adiabatic process, the adiabatic exponent is obtained,
\begin{equation}
\gamma=\theta+1,
 \label{gaa1}\end{equation}
which bridges the trap exponent $\theta$ and the adiabatic exponent
$\gamma$.  As an example, the trap exponent $\theta=2$ for a
one-dimensional box trap, and thus the adiabatic exponent $\gamma=3$
in this case, confirming the result obtained previously in a
different way \cite{Quan09}. The relation between the trap and
adiabatic exponents given by Eq. (\ref{gaa1}) can also be derived
very easily even if temperature is included \cite{note3}.

\begin{figure}[h]
\centering
\includegraphics[width=100pt]{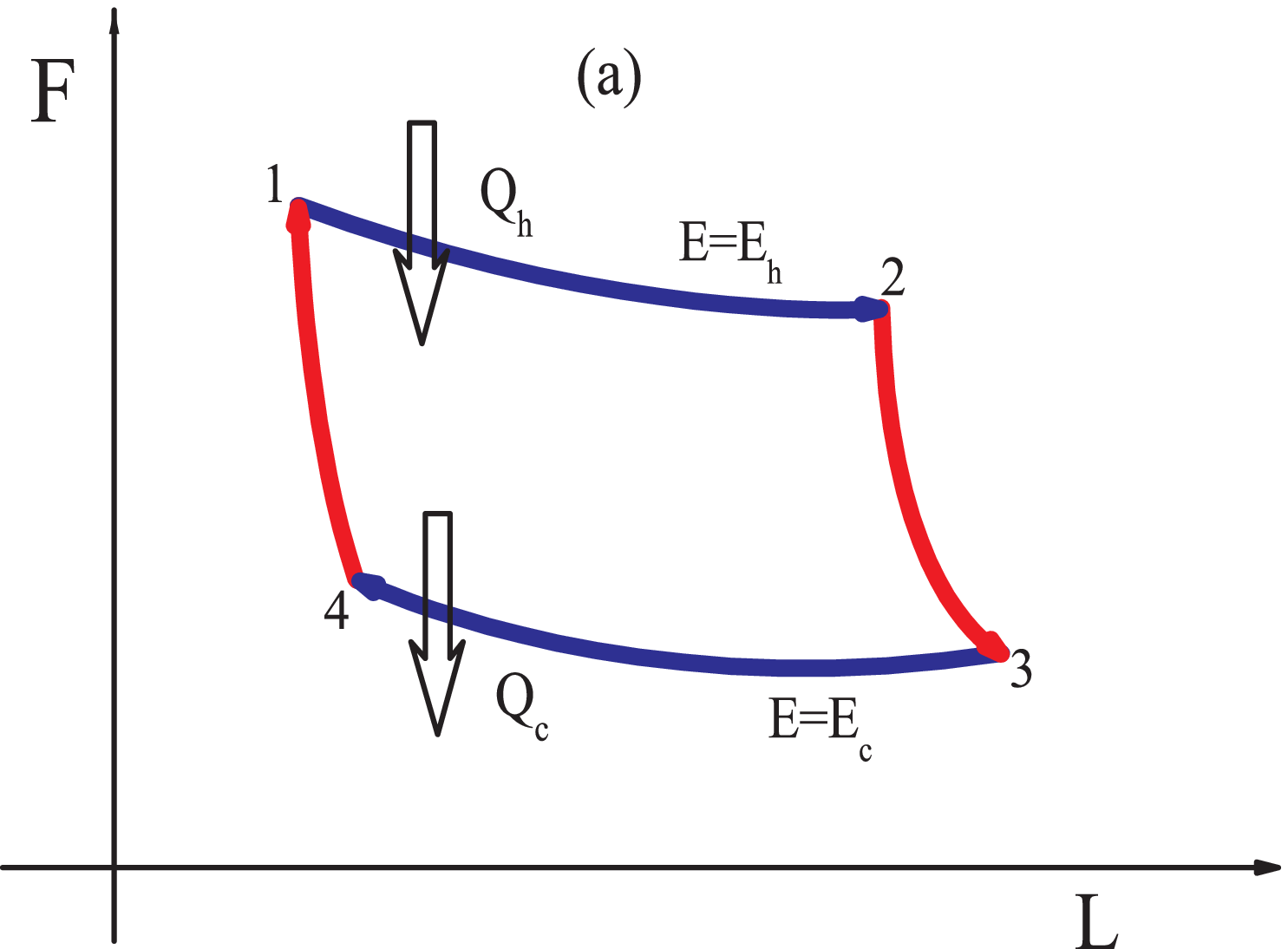}
\includegraphics[width=100pt]{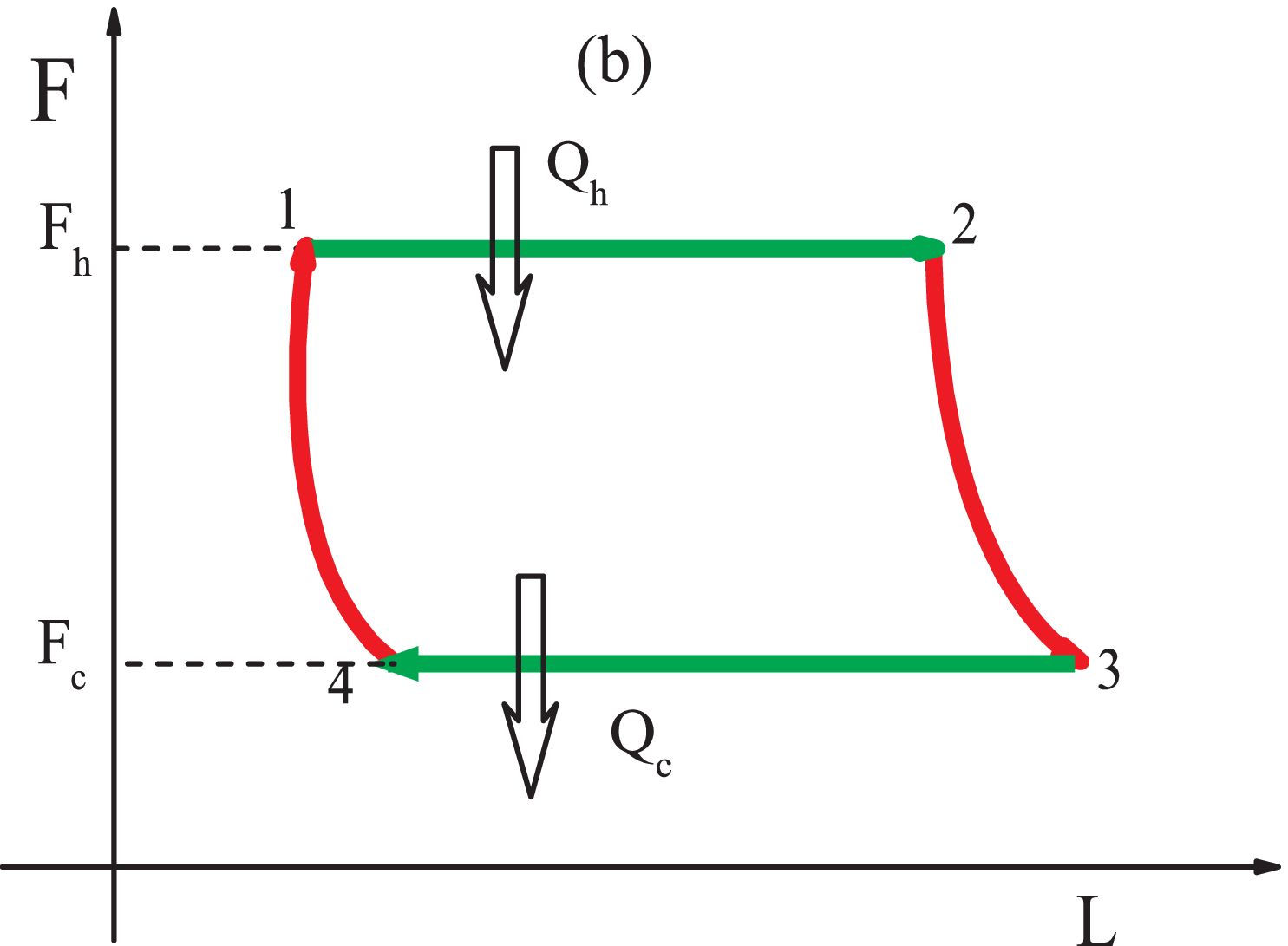}
\includegraphics[width=100pt]{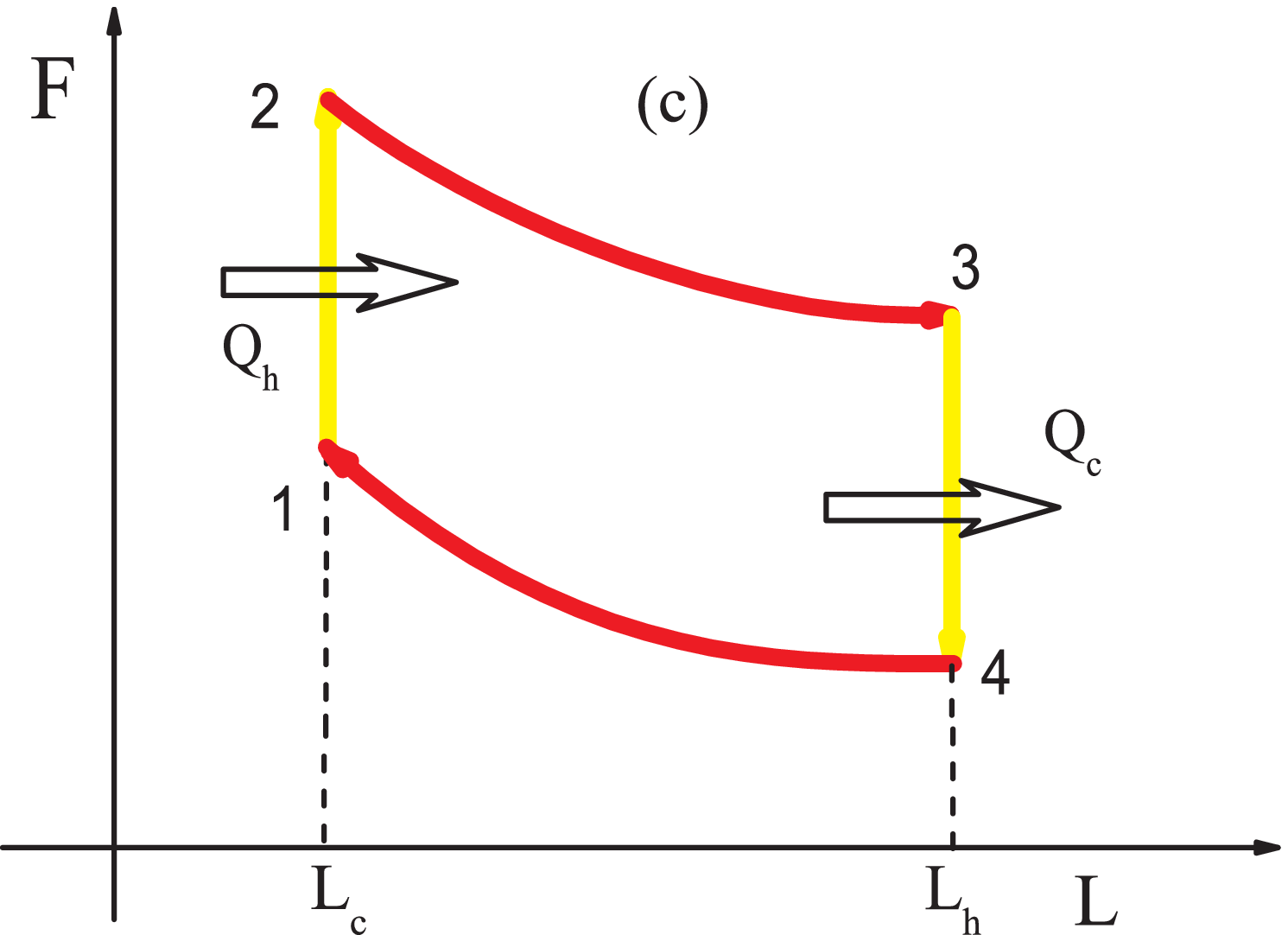}
\includegraphics[width=100pt]{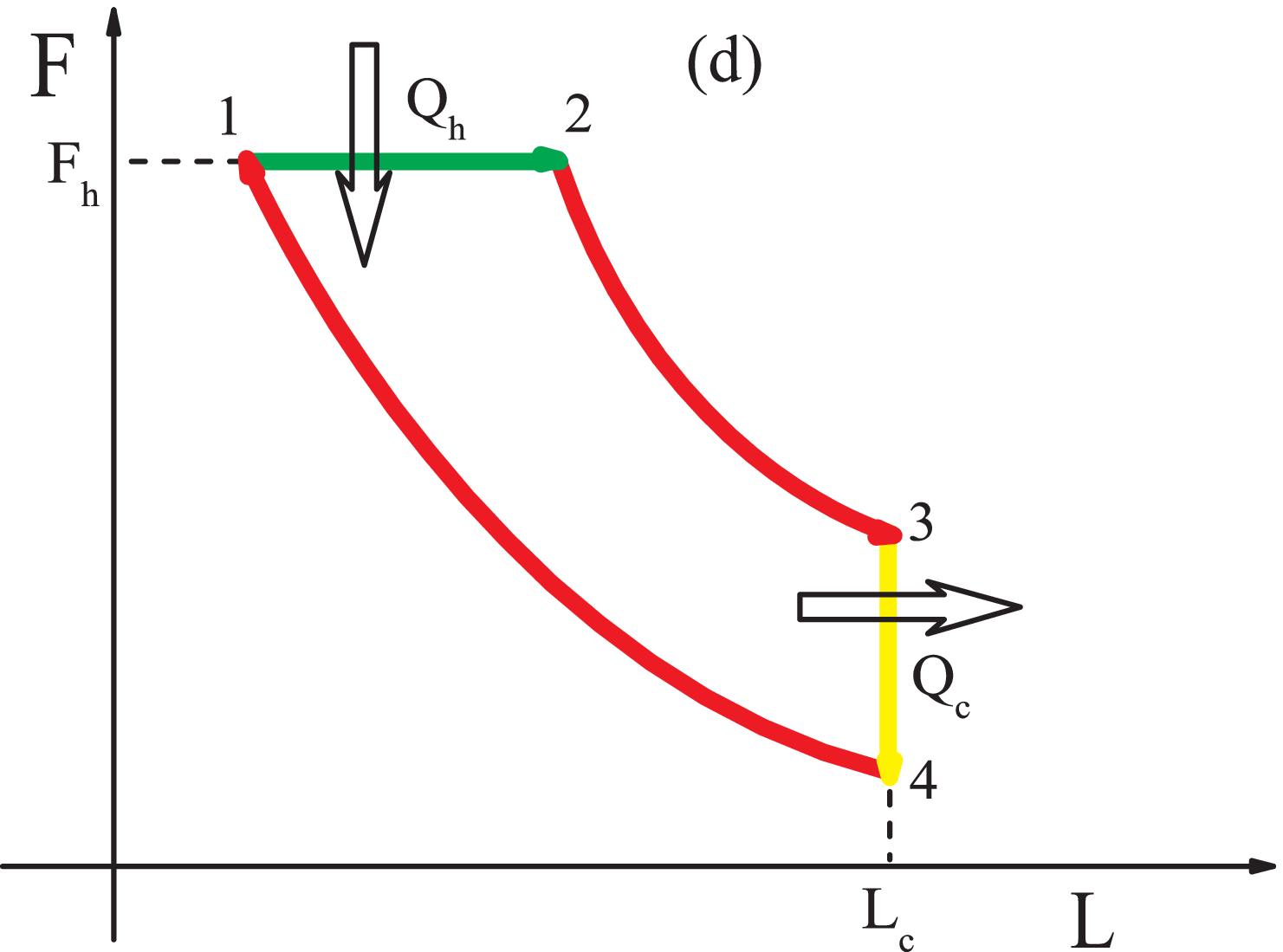}
\includegraphics[width=100pt]{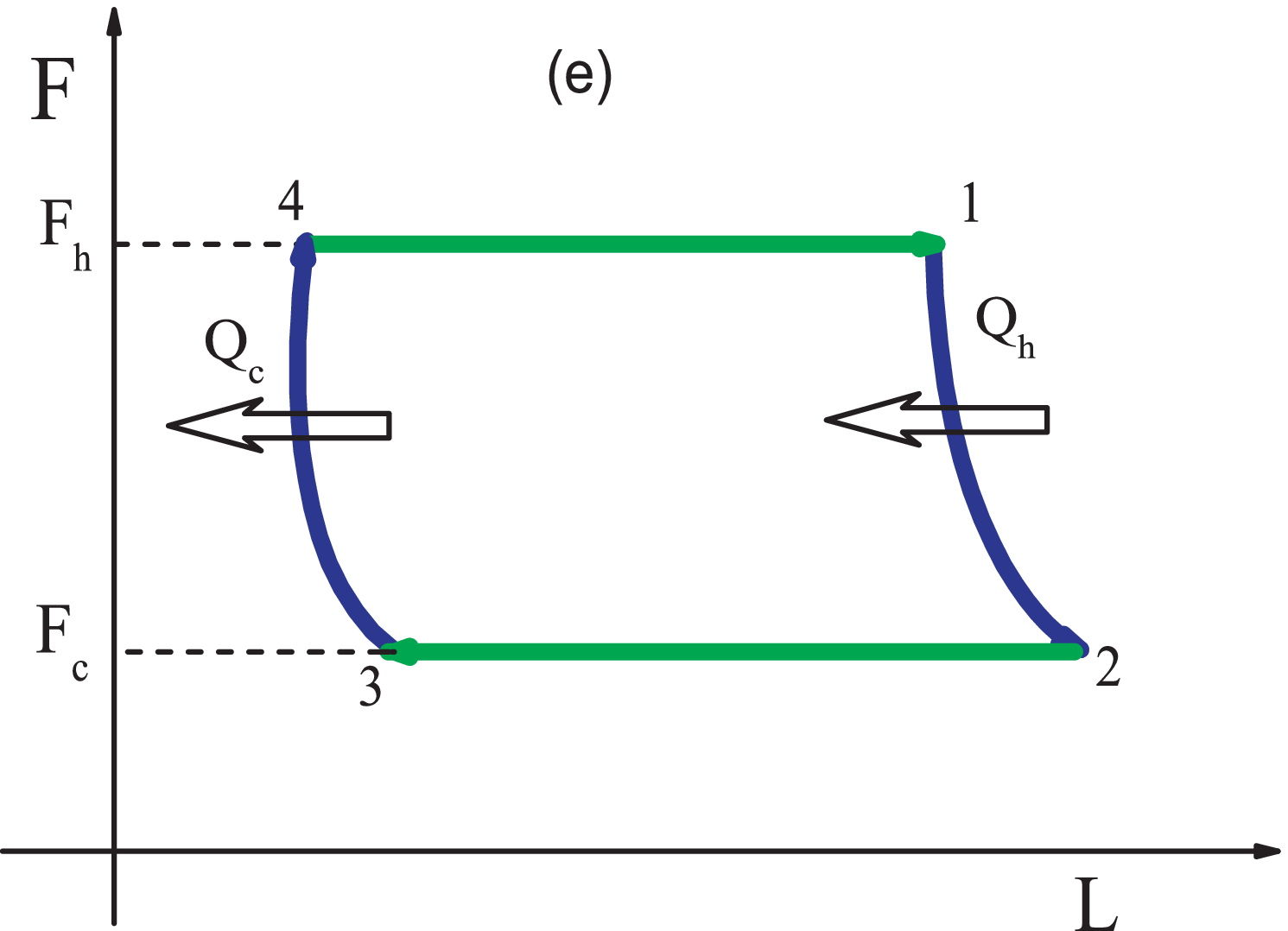}
\includegraphics[width=100pt]{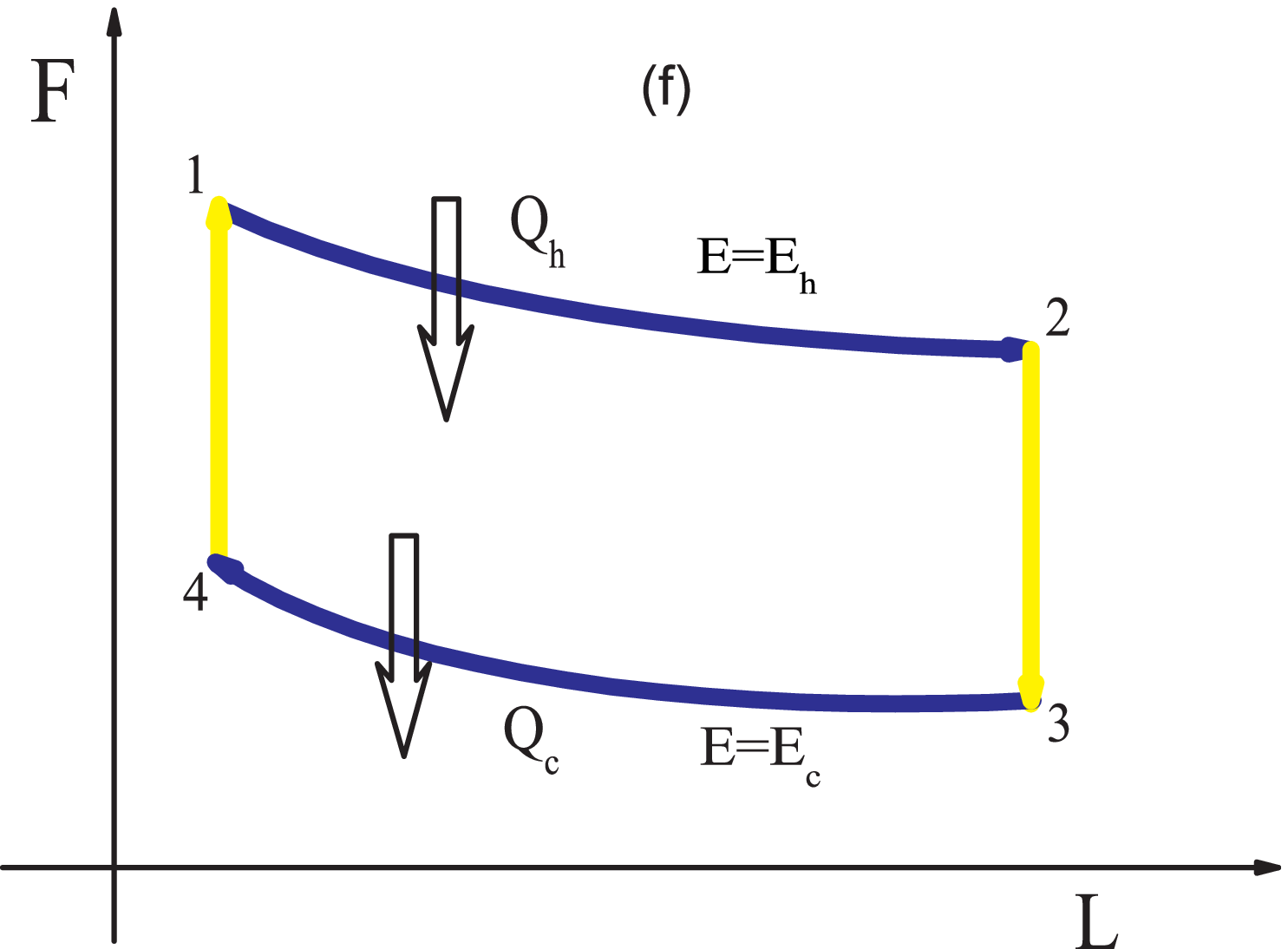}
\caption{(Color online) Graphic diagrams of QM (a) Carnot, (b)
Brayton, (c) Otto, (d) Diesel, (e) Ericsson, and (f) Stirling
cycles, in the plane of $(L, F)$. The isoenergetic, adiabatic,
isobaric, and isochoric processes, are indicated by the blue, red,
green, and yellow lines.} \label{fl}
\end{figure}

\emph{ {Various QM Engines And Their Efficiencies.}}
It is well known that any device, such as a heat engine,  or a fuel
cell, is described by its efficiency: the relationship between the
total energy input, and the amount of energy used to produce useful
work. To describe the performance of a QM engine, we follow this
definition of the efficiency:  the amount of energy input that is
actually converted to useful output.
 A QM Carnot cycle is
a QM analog of the classical as well as conventional quantum  Carnot
cycle, which consists of two quantum isoenergetic processes and two
quantum adiabatic processes, as shown in Fig. \ref{fl}(a). The
efficiency of the quantum Carnot cycle is given by
\begin{equation}
\eta_{C}=1-\frac{Q_{34}}{Q_{12}}=1-
\frac{E_{34}\ln{(L_3/L_4)}}{E_{12}\ln{(L_2/L_1)}},
 \label{eta21}
\end{equation}
where we have used Eq. (\ref{qili}).   Defining $E_h =E_{12}$, and
$E_c=E_{34}$, using Eq. (\ref{uifi}), the two constant energies of
the system $E_h$ and $E_c$  can be expressed as $E_h=F_1
L_1/\theta=F_2 L_2/\theta$ and $E_c=F_3 L_3/\theta=F_4 L_4/\theta$.
These two equations together with
$L_2^{\theta+1}F_{2}=L_3^{\theta+1}F_{3}$ and
$L_1^{\theta+1}F_{1}=L_4^{\theta+1}F_{4}$ obtained from Eq.
(\ref{ltns}), give rise to the relation $L_2/L_1=L_3/L_4$. Hence,
the efficiency of the QM Carnot engine becomes
\begin{equation}
\eta_{C}=1-\frac{E_c}{E_h}.
 \label{etuh}
\end{equation}
This result obtained earlier \cite{Ben00, Abe12, pre12III}, is the
same as that of a classical as well as quantum Carnot cycle working
between two heat reservoirs, with the identification of the system
energy as the temperature of the system, but it is derived here in a
general way.

A  QM Brayton cycle, consisting of two quantum isobaric and two
quantum adiabatic processes, is illustrated in Fig \ref{fl}(b).
Using Eq. (\ref{qiif}) and the fact that no heat exchange occurs in
two adiabatic processes, it follows that the efficiency of a quantum
Brayton cycle is
\begin{equation}
\eta_{B}=1-\frac{Q_{34}}{Q_{12}}=1-
\frac{(L_3-L_4)F_{34}}{(L_2-L_1)F_{12}}.
 \label{eteta}
\end{equation}
For two adiabatic processes $2\rightarrow 3$ and $4\rightarrow 1$,
from Eq. (\ref{ltns}) we have
$L_2^{\theta+1}F_{12}=L_3^{\theta+1}F_{34}$ and
$L_1^{\theta+1}F_{12}=L_4^{\theta+1}F_{34}$. Subtracting both sides
of these two equations, we can find,
\begin{equation}
\frac{L_2-L_1}{L_3-L_4}={\left(\frac{F_{34}}{F_{12}}\right)}^{\frac{1}{\theta+1}}.
\label{l2a1}
\end{equation}
Substitution of Eq. (\ref{l2a1}) into Eq. (\ref{eteta})  leads to
\begin{equation}
\eta_{B}=1-\left(\frac{F_{34}}{F_{12}}\right)^{1-1/\gamma}.
 \label{eta1}
\end{equation}
In deriving Eq. (\ref{eta1}),  the relation between $\theta$ and
$\gamma$ given by Eq. (\ref{gaa1}) has been used. This efficiency of
the QM Brayton engine is identical to that of a classical Brayton
cycle \cite{Per98}.

The engines operating by Otto cycle have been widely  used in
automobiles as well as the internal combustion engines
\cite{Quan07}. A  QM Otto cycle consisting of two isochoric and two
adiabatic processes is illustrated in Fig. \ref{fl}(c).  The
efficiency of the quantum Otto cycle can be expressed in terms of
pressures and potential widths at special instants,
\begin{equation}
\eta_{O}=1-\frac{L_3 F_3-L_4 F_4}{L_2 F_{2}-L_1 F_1}.
 \label{etf1}
\end{equation}
where $F_i$ and $L_i$, with $i=1, 2, 3, 4$, are pressures and
potential widths at four instants $1, 2, 3, $ and $4$. By denoting
$L_c=L_1=L_2$ and $L_h=L_3=L_4$ and using Eq. (\ref{ltns}), we
further simplifies Eq. (\ref{etf1}) to
\begin{equation}
\eta_{O}=1-\left(\frac{L_c}{L_h}\right)^{\theta}=1-\left(\frac{L_c}{L_h}\right)^{\gamma-1},
 \label{etao}
\end{equation}
which is the same as the efficiency of the classical Otto engine.

Besides two quantum adiabatic processes $2\rightarrow 3$ and
$4\rightarrow1$, the Diesel cycle consists of an isobaric process
$1\rightarrow 2$ and an isochoric process $3\rightarrow 4$ [see Fig.
\ref{fl}(d)]. The efficiency of the quantum Diesel cycle can be
expressed in terms of pressures and potential widths at special
instants,
\begin{equation}
\eta_{D}=1-\frac{\theta(E_3-E_4)}{(\theta+1)(L_2-L_1)F_h},
 \label{etfh}
\end{equation}
where we have taken $F_h=F_1=F_2$. Using
$F_hL_1^{\theta+1}=F_4L_3^{\theta+1}$ and
$F_hL_2^{\theta+1}=F_3L_3^{\theta+1}$, and $E_i=F_i L_i/\theta$,
with $i=3, 4$, we obtain $
\eta_{D}=1-\frac{1}{\theta+1}\frac{L_2^{\theta+1}-L_1^{\theta+1}}{(L_2-L_1)
L_3^{\theta}}, $ or
\begin{equation}
\eta_{D}=1-\frac{1}{\gamma}\frac{\left(\frac{L_2}{L_c}\right)^{\gamma}-\left(\frac{L_1}{L_c}\right)^{\gamma}}
{\left(\frac{L_2}{L_c}\right)-\left(\frac{L_1}{L_c}\right)},
 \label{etht}
\end{equation}
which is identical to its classical counterpart. In deriving Eq.
(\ref{etht}) we have used Eq. (\ref{gaa1}) and $L_c\equiv L_3$.


 A QM Ericsson (Stirling) cycle consists of two
quantum isoenergetic and two quantum isobaric (isochoric) processes.
The schematic diagram of the QM Ericsson as well as Stirling cycle
is plotted in Fig. \ref{fl}. From Eq. (\ref{uifi}), we find, in two
isobaric processes of the Ericsson cycle
 [see Fig.
\ref{fl}(e)], $L_1F_{41} = L_2F_{23}$ and $ L_3F_{23}= L_4F_{41}$,
where use of $F_1=F_4=F_{41}$ and $F_2=F_3=F_{23}$ has been made.
Hence, the amount of  energy  transported from the system,
$Q_{23}=(L_3-L_2)F_{23}$, is equal to that of  energy  absorbed by
the system, $Q_{41}=(L_4-L_1)F_{41}$. The efficiency of the QM
Ericsson engine is still given by the QM version of the Carnot
efficiency, $\eta_{E}=\eta_{C}=1-E_c/E_h$. In view of the fact that
no work is done in an isochoric process, we find that the  energy
absorbed by the system during the isochoric process $2\rightarrow
3$, $Q_{23}=E_h-E_c$, is totally counterbalanced by the energy
released to the system in the isochoric process $4\rightarrow 1$,
$Q_{41}=E_h-E_c$ [see Fig. \ref{fl}(f)]. The expression of
efficiency for  the QM Stirling engine is thus the same as that for
the QM Carnot as well as Ericsson engine, namely,
$\eta_S=1-{E_c}/{E_h}$.

Before ending this section, we would like to emphasize that the
energy spectrum as well as the occupation probabilities considered
here are general and intrinsic. Therefore, our result in the present
paper is independent of any parameter contained in the system,
whether the interaction between particles is considered or not, and
it is indeed valid for an arbitrary ideal or interacting system,
such as a system with an arbitrary number of particles in an
arbitrary power-law trap, a harmonic system,  a spin-$1/2$ system,
and a single-mode radiation field, and so on.

\emph{Conclusion.} We have studied the  energy  analogy of the
classical thermodynamic cycles based on  microscopic definitions of
various thermodynamic processes. We have clarified the properties of
these quantum thermodynamic processes and cycles, bridging the
quantum thermodynamic cycles and their classical counterparts.
Comparison between quantum adiabatic process and its classical
counterpart gives rise to a relation between the trap exponent and
the quantum adiabatic exponent. The universality of the efficiency
for any given cycle is identified, in the sense that the expression
of the efficiency is intrinsic and independent of any parameter
involved in a given engine model.

\emph{Acknowledgements:} We gratefully acknowledge support for this
work by the National Natural Science Foundation of China under
Grants No. 11265010, No. 11147200, No. 11065008, No. 10974033, and
No. 11191240252, the State Key Programs of China under Grant No.
2012CB921604, and the Foundation of Jiangxi Educational Committee
under Grant No. GJJ12136. J. H. Wang is very grateful to Haitao Quan
for his valuable discussions.


\begin{thebibliography}{99}
\bibitem{Gem09} J. Gemmer, M. Michel, and G. Mahler, Quantum Thermodynamics,
2nd ed. (Springer-Verlag, Berlin, 2009).
\bibitem{Per98} P. Perrot, \emph{A to Z of Thermodynamics }(Oxford University Press,
Oxford, 1998).

\bibitem{Gev92} E. Geva and R. Kosloff, J. Chem. Phys. \textbf{96}, 3054 (1992);
 J. Chem. Phys. \textbf{97}, 4396(1992); J. Chem. Phys. \textbf{102}, 8541 (1995);
 R. Kosloff, E. Geva, and J. Gordon, J. Appl. Phys. \textbf{87}, 8093 (2000).

\bibitem{Scu03} M. O. Scully, M. S. Zubairy, G. S. Agarwal, and H. Walther, Science
\textbf{299}, 862 (2003).

\bibitem{Kim11} S. W. Kim, T. Sagawa, S. De Liberato, and M. Ueda, Phys. Rev. Lett. \textbf{106}, 070401
(2011).

\bibitem{Fia12} O. Fialko  and D. W. Hallwood, Phys. Rev. Lett. \textbf{108}, 085303 (2012).

\bibitem{Quan07} H. T. Quan, Y. X. Liu, C. P. Sun, and F. Nori,
Phys. Rev. E \textbf{76}, 031105 (2007).

\bibitem{Quan09} H. T. Quan, Phys. Rev. E \textbf{79}, 041129 (2009).


\bibitem{Ben00} C. M. Bender,  D. C. Brody, and B. K. Meister, J. Phys. A: Math. Gen. \textbf{33}, 4427 (2000).

\bibitem{Abe12} S. Abe, Phys. Rev. E \textbf{83}, 041117 (2011); S. Abe and S. Okuyama, Phys. Rev. E \textbf{83}, 021121
(2011); S. Abe  and S. Okuyama, Phys. Rev. E \textbf{85}, 011104
(2012).

\bibitem{pre1203} J. H. Wang, J. Z. He, and Z. Q. Wu, Phys. Rev. E \textbf{85}, 031145 (2012).

\bibitem{pre1104} J. H. Wang, J. Z. He, and X. He, Phys. Rev. E \textbf{84}, 041127 (2011).

\bibitem{pre12III} R. Wang, J. H. Wang, J. Z. He, and Y. L. Ma, Phys. Rev. E \textbf{86}, 021133 (2012);
 J. H. Wang and J. Z. He, J. Appl. Phys. \textbf{11}, 043505 (2012).

\bibitem{pre1204} J. H. Wang, Z. Q. Wu, and J. Z. He, Phys. Rev. E \textbf{85}, 041148
 (2012).

\bibitem{Lu12} Y. Lu and G. L. Long, Phys. Rev. E \textbf{85}, 011125 (2012).

\bibitem {Fel00} T. Feldmann and  R. Kosloff, Phys. Rev. E \textbf{61}, 4774 (2000);
T. Feldmann and R. Kosloff, Phys. Rev. E \textbf{68}, 016101 (2003);
T. Feldmann and R. Kosloff, Phys. Rev. E \textbf{70}, 046110 (2004).

\bibitem{Wu06} F. Wu, L. G. Chen, S. Wu,
F. R. Sun, and  C. Wu, J. Chem. Phys. \textbf{124}, 214702 (2006);
F. Wu, L. G. Chen, F. R. Sun, C. Wu, and Q. Li, Phys. Rev. E
\textbf{73}, 016103 (2006).

\bibitem{Rez06} Y. Rezek and R. Kosloff, New J. Phys. \textbf{8}, 83 (2006).

\bibitem{Huang12} X. L. Huang,  L. C. Wang,  and X. X.
Yi,  arXiv:1209.1684 [quant-ph].
\bibitem{He02} J. Z. He, J. C. Chen, and  B. Hua, Phys. Rev. E \textbf{65}, 036145 (2002).

\bibitem{Cle12} B. Cleuren, B. Rutten, and C. Van den Broeck, Phys. Rev. Lett.
\textbf{108}, 120603 (2012).

\bibitem{Kie04} T. D. Kieu, Phys. Rev. Lett.,  \textbf{93},  140403 (2004).
\bibitem{All08} A. E. Allahverdyan, R. S. Mahler, and G. Johal,  Phys. Rev. E, \textbf{77},
041118(2008).
 \bibitem{Hul12} X. L. Huang, T. Wang, and X. X. Yi, Phys. Rev. E \textbf{86},
051105 (2012).
\bibitem{Quan05} H. T. Quan, P. Zhang, and C. P. Sun, Phys. Rev. E \textbf{72}, 056110 (2005).

\bibitem{Hui11} J. H. Wang, J. Z. He, and Y. L. Ma, Phys. Rev. E   \textbf{83}, 051132 (2011);
H. Y. Tang  and Y. L. Ma, Phys. Rev. E \textbf{83}, 061135 (2011).
\bibitem{Ben02} C. M. Bender, D. C. Brody, and B. K. Meister, Proc. R. Soc. Lond. A
\textbf{458}, 1519 (2002).



\bibitem{epjd} J. H. Wang and J. Z. He, Eur. Phys. J. D \textbf{64}, 73
(2011); J. Low. Temp. Phys. \textbf{166}, 80 (2012).



\bibitem{Wil97} M. Wilkens and C. Weiss, J. Mod. Opt. \textbf{44}, 1801 (1997); C.
Weiss and M. Wilkens, Opt. Express \textbf{1}, 272 (1997).




\bibitem{wpra09}J. H. Wang, H. Y. Tang, and Y. L. Ma, Ann. Phys.
\textbf{326}, 634 (2011), and references therein.

\bibitem{Lin76} G. Lindblad, Commun. Math. Phys. \textbf{48}, 119 (1976); U. Weiss,
\emph{Quantum Dissipative Systems}, 2nd ed. (World Scientific,
Singapore, 1999).

\bibitem{Foc28} M. Born and V. Fock, Z. Phys. \textbf{51}, 165 (1928).



\bibitem{arx12} J. H. Wang and J. Z. He, Phys. Rev. E \textbf{86}, 051112
(2012).
\bibitem{note3} In the presence of a heat bath,  the relation between the trap and adiabatic
exponents can be readily obtained without requiring derivation of an
explicit function of the entropy $S$.
  When a $N$-particle quantum system is in thermal equilibrium with
a heat reservoir at constant temperature $T$, the occupation
probability $p_n$ at any state $n$  of the system must satisfy the
Boltzmann distribution: $p_n=\frac{1} {Z_N} e^{-\varepsilon_n/{(k_B
T)}}$, where $k_B$ is the Boltzmann constant and
$Z_N=\sum_ne^{-\varepsilon_n/{(k_B T)}}$ is the canonical partition
function.   For the quantum adiabatic process the probabilities
$p_n$ remain  constant thus leading to constant entropy since $S$ is
determined by $S=-\sum_n k_B p_n\ln p_n$. Whereas the temperature
$T$ varies in an adiabatic process, the ratio $\varepsilon_n/T$
should keep fixed in order for $p_n$ which is only the function of
$\varepsilon_n/T$ to be a constant. Adopting the energy spectrum
given by Eq. (\ref{enma}), we find that $T L^\theta=\texttt{cons}$,
which yields the relation $\theta=\gamma-1$ through comparison with
 $TL^{\gamma-1}=\texttt{cons}$  for the adiabatic
process.

\bibitem{n2}  For the convetional Carnot cycle, we find that the Carnot
efficiency, $\eta_C=1-Q_c/Q_h=1-T_c/T_h$, and that
$\eta_E=\eta_S=1-T_c/T_h$. Since the heat exchange in an isobaric
(isochoric) process, in which the system couples to a heat bath
instead of an energy bath, is still given by Eq. (\ref{qiif})[Eq.
(\ref{qiui})],  the expressions of the effiencies for the ideal
Brayton, Otto, and Diesel cycles are still given by Eqs.
(\ref{eta1}), (\ref{etao}), and (\ref{etht}), respecively.




\bibitem{Bli12} V. Blickle and C. Bechinger, Nat. Phys. \textbf{8}, 143 (2012).


\end{thebibliography}
\end{document}